\documentclass[preprint,amsmath,amssymb,aps,prd,nofootinbib]{revtex4}
\usepackage{graphicx}

\begin{document}

\title{Remarks on the lifetime of sterile neutrinos and \\
the effect on detection of rare meson decays $M^+\to M^{\prime -} \ell^+\ell^+$}

\thispagestyle{empty}

\author{Claudio Dib$^1$\footnote{claudio.dib@usm.cl}, C. S. Kim$^2$\footnote{Corresponding author, ~cskim@yonsei.ac.kr}}
\affiliation{ $^1$ Dept. of Physics and CCTVal, Universidad T. Federico Santa Mar\'\i a, Valpara\'\i so, Chile \\
$^2$ Dept. of Physics and IPAP, Yonsei University, Seoul 120-749, Korea
}

\begin{abstract}
\noindent
We review the detection of the $\Delta L=2$ semileptonic meson decays of the form $M^+\to
M^{\prime -} \ell^+\ell^+$ that we studied in a previous work, mediated by a Majorana neutrino with a mass in the range between the masses of the initial and final mesons. For such range of masses, the Majorana neutrino will go on its mass shell and the two charged leptons will appear at displaced vertices or, if the lifetime is long enough, most secondary vertices will fall outside the detector. We study the consequences of this effect on the experimental searches and limits that can be extracted.

%\noindent Keywords :

\end{abstract}
 \maketitle \thispagestyle{empty}

Here we briefly comment on the search for sterile  Majorana neutrinos through $\Delta L=2$ meson decays
\cite{DeltaL}, especially
the decays of charged $K,~ D,~ D_s,~ B$, and $B_c$ mesons of the form
$M^+ \to M'^- l^+ l^+$, induced by the existence of Majorana neutrinos, as shown in Fig. 1 and studied in a previous work \cite{BDK}.
\begin{figure}[h] %%%%%%%%%%%%%%%% FIGURE 1 %%%%%%%%%%%%%%%%%%%%%%%%%%%%%%%%%%%
\begin{minipage}[b]{.49\linewidth}
 \includegraphics[width=\linewidth]{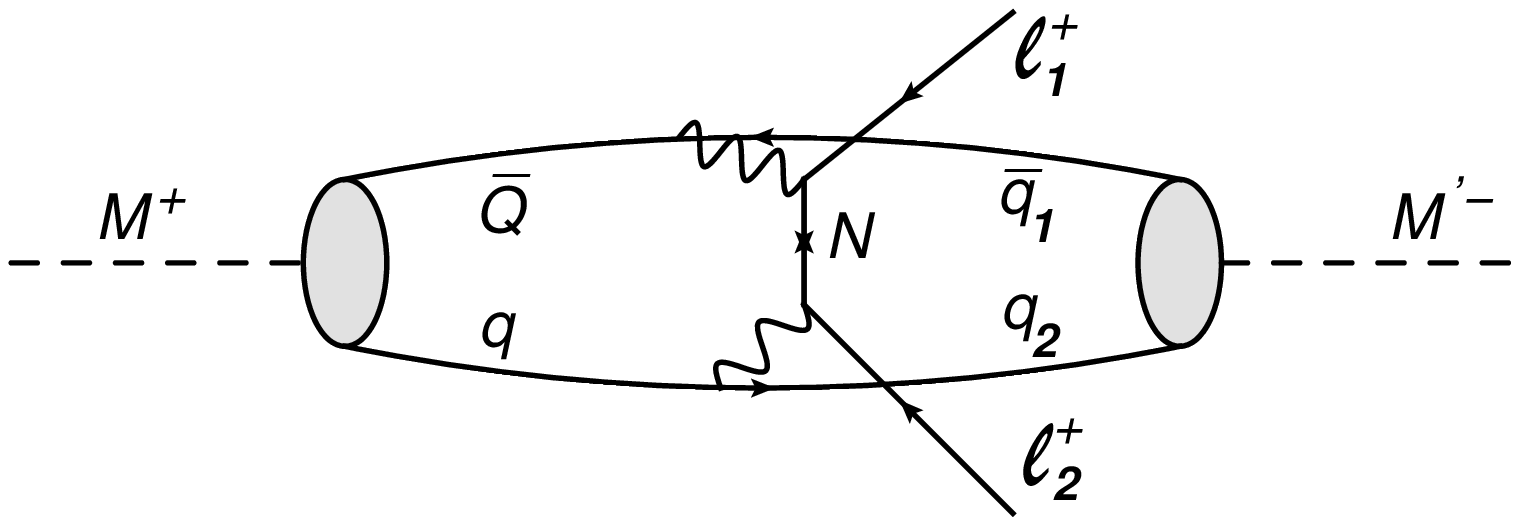}
\end{minipage}
\begin{minipage}[b]{.49\linewidth}
\vspace{0pt}
 \includegraphics[width=\linewidth]{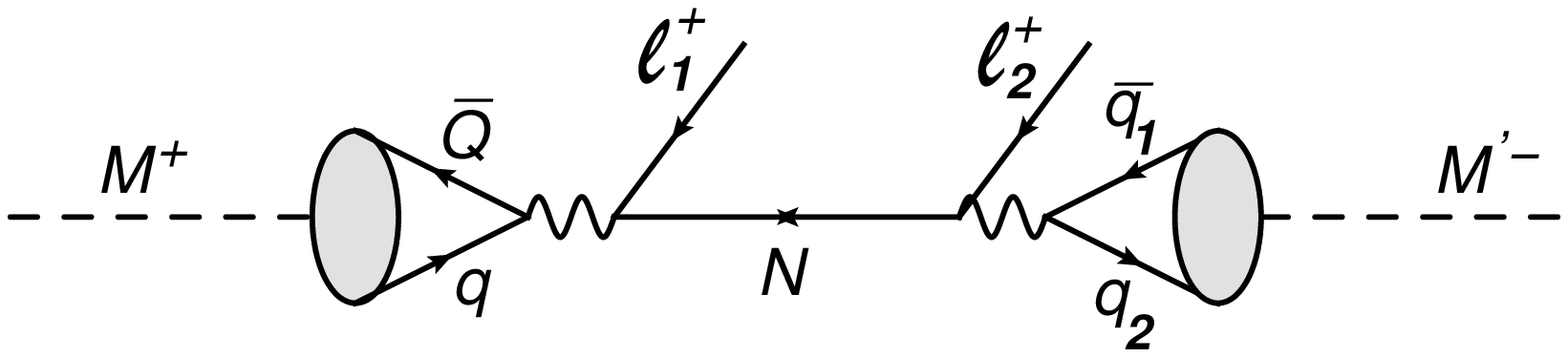}
\end{minipage}
\vspace{-0.2cm}
\caption{The $t$-type and $s$-type  weak amplitudes at the quark
level that enter in the process $M^+\to {M'}^- \ell_1^+ \ell_2^+$
(plus the same diagrams with leptons exchanged if they are
identical). $M^+$ and $M'^-$ are pseudoscalar mesons, such as $B_c,~B,~D_s,~K$ and $\pi$.} \label{fig1}
\end{figure}

For a sterile neutrino with mass $m_N$ in the range $(m_{M'^-}, m_{M^+})$, the $s$-type diagram (Fig.~1.b) will dominate the process as the neutrino $N$ can go on its mass shell \cite{Ivanov}. Since the lifetime of a weakly interacting particle such as $N$ is rather long, in an experiment where the decay $M^+\to M^{\prime -} \ell^+\ell^+$ is produced, the spatial position of the vertices where the two charged leptons are produced will be displaced.
 Therefore, the search for a massive neutrino $N$ in these processes should first be done by looking at the energy spectrum of the positive lepton
 produced at the primary vertex, $i.e.$ in a search for the $M^+$ decay mode $M^+ \to \ell^+ N$.  The partial decay rate of this mode, neglecting the mass of the charged lepton, is given by:

\begin{equation}
\Gamma(M^+\to \ell^+ N) = \frac{1}{8\pi} G_F^2 f_M^2 |V_{Qq}|^2 m_M  m_N^2    \left(1-
\frac{m_N^2}{m_M^2}\right)^2
|U_{\ell N}|^2,
\label{Mdecaymode}
\end{equation}
where $f_M$ is the decay constant of the decaying meson, $V_{Qq}$ is the CKM quark mixing element associated to the meson decay, and $U_{\ell N}$ is the lepton mixing element of the massive neutrino $N$ with a standard lepton flavor $\ell$.

Now, given that the total width $\Gamma_N$ of the intermediate Majorana neutrino $N$ will clearly be much smaller than its mass, as will be shown in Fig. 2,
one can use the narrow width approximation to express the decay dominated by Fig. 1(b) as the decay rate of the primary process times the branching ratio of the $N$ decay subprocess, namely:
\begin{equation}
\Gamma(M^+\to \ell^+ \ell^+ M'^-) =  \Gamma(M^+\to \ell^+ N)\cdot Br(N \to \ell^+ M'^-),
\end{equation}
where  $Br(N\to \ell^+ M'^-) \equiv \Gamma(N\to \ell^+M'^-)/\Gamma_N$,  and
\begin{equation}
\Gamma(N \to \ell^+ M'^-) = \frac{1}{16\pi} G_F^2 f_{M'}^2 |V_{qq'}|^2 m_N^3 \left(1 -
\frac{m_{M'}^2} {m_N^2}\right)^2
|U_{\ell N}|^2.
\label{Ndecaymode}
\end{equation}
Here one should notice that the total decay width of $N$, $\Gamma_N$, is proportional to $m_N^5 |U_{\ell N}|^2$ (summed over all $\ell$).
Hence, the branching fraction, $Br(N\to \ell^+ M'^-)$, is rather independent of the mixing value $U_{\ell N}$.

Theoretically, the branching ratio for the $M^+$ decay, with unlimited detectability of the full final state, would simply be
\begin{equation}
Br(M^+\to \ell^+ \ell^+ M'^-)_{\rm th} =  Br(M^+\to \ell^+ N)\cdot  Br(N \to \ell^+ M'^-).
\end{equation}
However,  for a large lifetime of $N$, not all decays $N\to \ell^+ M^{\prime -}$ will occur inside a detector of finite size. The probability for the neutrino $N$ to decay inside a detector with length $L_D$ from the primary vertex is (with $\hbar = c = 1$):
%
%\begin{equation}
%P_N  = 1- \exp\left(-\frac{L_D}{\tau_N\gamma_N\beta_N}\right) = 1- \exp\left(-\frac{\Gamma_N L_D}{\gamma_N\beta_N}\right),
%\label{PN}
%\end{equation}
%
\begin{equation}
P_N  = 1- e^{-L_D/L_N} ,
\label{PN}
\end{equation}
where $L_N =  \gamma_N \beta_N \tau_N $ is the ``decay length'' of the neutrino $N$,
being $\beta_N$ and $\gamma_N$ its velocity and relativistic factor, respectively. For neutrinos of very long lifetimes and highly relativistic motion,
\hbox{$P_N \approx  L_D/L_N \approx L_D/\tau_N \gamma_N  \ll 1$}.
Considering a general $P_N$, the experimentally observed branching ratio actually is
\begin{equation}
Br(M^+\to \ell^+ \ell^+ M'^-)_{\rm ex} =  Br(M^+\to \ell^+ N)\times  P_N \times  Br(N \to \ell^+ M'^-).
\end{equation}
As such, $P_N$ is an acceptance factor, so that  the correct branching ratio, $Br_{\rm th}$, can be estimated from the experimentally observed ratio $Br_{\rm ex}$ after multiplying the latter by $1/P_N$.  If an experiment only gives the upper limit, the limit would be weakened by the factor $1/P_N$.

\begin{figure}[ht] %%%%%%%%%%%%%%%%% FIGURE 2 %%%%%%%%%%%%%%%%%%%%%%
  \centering
  \includegraphics[scale=1]{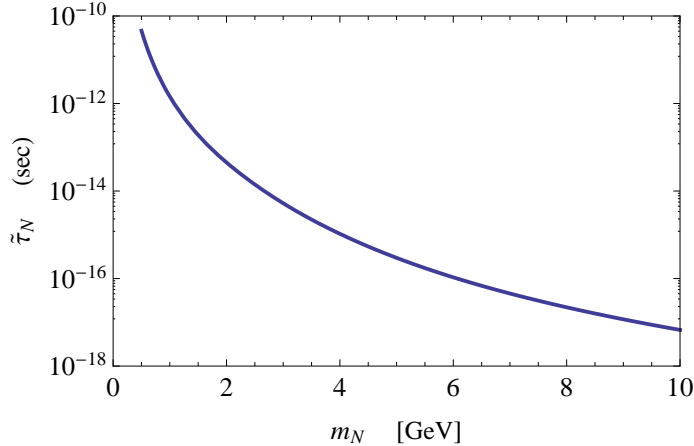}
  \caption{The normalized $N$ neutrino lifetime, $\tilde\tau_N$ (using lepton mixing $|U_{\ell N}|^2=1$) vs. its mass $m_N$. Actual lifetime is $\tau_N = \tilde\tau_N/|U_{\ell N}|^2$.  $\tilde\tau_N$ is estimated as inclusive decay into all allowed standard quarks and leptons and  assumes equal $|U_{\ell N}|^2$ for all three flavours $\ell = e, \mu, \tau$. Current bounds are $|U_{\ell N}|^2< 10^{-9}$ for $m_N\lesssim 500$ MeV, and  $|U_{\ell N}|^2< 10^{-7}$ for $m_N\gtrsim 1$ GeV \cite{Atre}.}
%
%  \caption{The normalized lifetime $\tilde\tau_N$ of the sterile neutrino $N$ as a function of its mass, $m_N$, using the same lepton mixing  $|U_{\ell N}|^2=10^{-9}$ for all three flavours $\ell = e, \mu, \tau$. Since $\tau_N\propto 1/|U_{N\ell}|^2$, for other values of the mixing the lifetime can be scaled accordingly. The lifetime of $N$ is estimated by assuming inclusive decays into all kinematically allowed standard quarks and leptons.}
  \label{fig2}
\end{figure}

Presently the experiments, CLEO \cite{CLEO}, Belle \cite{Belle}, BaBar \cite{BaBar} and LHC-b \cite{LHC-b}
have used $P_N =1$ in their searches for the $M^+\to M^{\prime -} \ell^+\ell^+$ decays, which is equivalent to assume that the sterile neutrino $N$ has a lifetime short enough that all events
decay at a single vertex, or at least decay fully inside the detector.
This is correct only for very heavy $N$ and large $U_{\ell N}$.
 Fig. 2  shows the possible lifetimes $\tau_N$ of the sterile neutrino $N$ as a function of its mass %$m_N$, assuming a mixing $|U_{\ell N}|^2=10^{-9}$
  \cite{Atre,Helo2,CPVpi}.
 For $m_N = 4$~GeV or less, the lifetime is $\tau_N \sim 10^{-8}$ sec or longer, which implies average vertex displacements of the order $L_N\sim 1$~meter or longer. Consequently, the effect of $P_N$ clearly has to be taken into account.

%
%This is correct only for relatively heavy $N$ and large $U_{\ell N}$.
% Fig. 2  shows the possible lifetimes $\tau_N$ of the sterile neutrino $N$ as a function of its mass $m_N$, assuming a mixing $|U_{\ell N}|^2=10^{-9}$ \cite{Atre,Helo2,CPVpi}. For $m_N = 6$~GeV or less, the lifetime is $\tau_N \sim 10^{-7}$ sec or longer, which implies average vertex displacements of the order $L_N \sim 10$~meters or longer. Consequently, the effect of $P_N$ clearly has to be taken into account.

Belle \cite{Belle} precisely described ``All charged tracks are required to originate near
the interaction point and have impact parameters within
5 cm along the beam direction and within 1 cm in the
transverse plane to the beam direction." With the KEKB asymmetric-energy $e^+ e^-$ collider (3.5 on 8 GeV),
$\gamma_N  \sim 2$, and let us  take $\tau_N \sim 10^{-7}$ sec according to Fig. 2, and therefore the decay length is $L_N\sim 10^2$~m.
As an example,
the measurement of Belle for  $Br(B^+ \to e^+ e^+ D^-)_{\rm ex}$ can be adjusted to
\begin{eqnarray}
Br(B^+ \to e^+ e^+ D^-)_{\rm th} &=& Br(B^+ \to e^+ e^+ D^-)_{\rm ex} / P_N \\
&\simeq& Br(B^+ \to e^+ e^+ D^-)_{\rm ex} \times [100\, m] /L_D ,
\end{eqnarray}
which implies the correct upper bound on the branching ratio would be about 100 times larger than the experimentally
measured value of the upper bound  in a detector length around  $L_D=1\ m$.

A similar  example is the $\Delta L=2$ kaon decay $K^+\to \pi^-\mu^+\mu^+$. Here the neutrino $N$  that can go on mass shell must have a mass $m_N$ in the range $250 - 400$ MeV. The corresponding lifetime of the neutral lepton, $\tau_N$, ranges between 60 ms to 300 ms, for $m_N$ 400 and 250 MeV, respectively (here we have used a conservative bound on the mixing $|U_{\ell N}|^2\sim 10^{-9}$; since
$\tau \propto 1/|U_{\ell N}|^2$, for other values of mixing $\tau_N$ can be scaled accordingly). For kaons not too relativistic, the intermediate neutrino $N$ will
be slightly relativistic, so we can estimate the decay length $L_N = \gamma_N \beta_N
\tau_N  c \sim \tau_N c$, which is  $2\times 10^7 - 1\times 10^8$ m. For a detector 100 m long,
the acceptance factor is then $P_N \sim 1 - 5 \times 10^{-6}$, for $m_N$ $250 - 400$ MeV, respectively.

We now briefly comment on the LHC experiments and on the possible $\pi$ decay experiments  from \hbox{Project-X} to probe Majorana neutrinos.
For the LHC, at which much heavier $N$
can be probed ($m_N \sim 100-1000$ GeV), $P_N$ could be almost $1$,  still depending on the value of $U_{\ell N}$. Actually the experiment can be used to constrain the limit
of $U_{\ell N}$ at those large masses \cite{Helo}.
On the other hand, for the $\Delta L=2$ pion decays $\pi^+ \to e^+ e^+ \mu^- \nu$
that could be observed at Project-X, the neutrinos that
can go on mass shell must have masses below $m_\pi$ \cite{pi}. These neutrinos would be very long-living: $\tau_N \gtrsim 10^{-8} /|U_{eN}|^2$ seconds,
which is 10 seconds or longer, if the mixing is near the conservative bound $|U_{eN}|^2 \sim 10^{-9}$.
The acceptance  in such cases would be
\[
P_N \sim \frac{|U_{eN}|^2}{\gamma_\pi} \left(\frac{L_D}{[m]}\right)
\]
or less. If we take the Lorentz factor expected for Project-X at $\gamma_\pi \sim 10$ and $|U_{eN}|^2 \sim 10^{-9}$, we should expect
acceptances $P_N\sim 10^{-10} (L_D/[m])$ or less.

As a final example, we may consider the new proposal to detect heavy neutrinos through charmed meson decays using the SPS at CERN \cite{Bonivento:2013jag}, where the decay length has been  taken into account.  In this case, $m_N\sim 1$ GeV is expected, which should have a lifetime $\tau_N \sim 10^{-5}$~seconds if we use  the bound $|U_{N\ell}|^2\sim 10^{-7}$ for such mass. For a detector length $L_D\sim 10^2$~meters as in the proposal, rather large acceptance factors,  $P_N\sim 10^{-1}/\gamma_N$, should be expected.
\\

To summarize, we point out that when studying $\Delta L=2$ meson decays of the form $M^+ \to M^{\prime -} \ell^+\ell^+$, which occur if there exist  heavy neutral leptons $N$ of Majorana type, these processes are by far dominated when $N$ goes on its mass shell in the $s$-channel, provided its mass is in the intermediate range $m_{M'}+m_\ell < m_N < m_M-m_\ell$. A neutral lepton with such mass must be sterile, so the process must occur via lepton mixing, and the experimental study of these processes can help put bounds on the mixing element $U_{\ell N}$. However, one should consider that a neutral lepton with masses below a few GeV is necessarily long living, and therefore the process will proceed in the sequence $M^+\to \ell^+ N$ followed by $N\to\ell^+ M'^-$, where the two leptons will appear at separate vertices. The vertex  displacement can be very large, more so for smaller masses, so that most secondary decays may occur outside the detector. In such cases, one should consider the correct acceptance factor at the moment of deducing upper bounds for the lepton mixing parameter as a function of $m_N$.
\\

\noindent {\bf Acknowledgements:}
\noindent
We thank  Youngjoon Kwon, G.~Cvetic  and J.~Zamora-Sa\'a for helpful discussions.
C.S.K. was supported by the National Research Foundation of Korea (NRF)
grant funded by Korea government of the Ministry of Education, Science and
Technology (MEST) (No. 2011-0017430) and (No. 2011-0020333). C.D. acknowledges support
from FONDECYT (Chile) grant 1130617.

\end{document}